\documentclass[english,english,pra,english,preprint,amsmath,amssymb,aps,prl,superscriptaddress,longbibliography,showkeys, titlepage]{revtex4-2}
\usepackage{lmodern}
\usepackage[T1]{fontenc}
\usepackage[latin9]{inputenc}
\usepackage{amsmath}
\usepackage{amsthm}
\usepackage{amssymb}
\usepackage{graphicx}

\makeatletter

\usepackage[T1]{fontenc}
\usepackage{amsmath}
\usepackage{amsthm}
\usepackage{amssymb}
\usepackage{graphicx}

\usepackage{graphicx}
\usepackage{float}
\usepackage{xcolor}
\usepackage[hypertexnames=false]{hyperref}
\hypersetup{
	breaklinks = true,
    colorlinks = true,
    citecolor = {blue},
	urlcolor = {blue},
	linkcolor = {blue}
}

\ifdefined\showcaptionsetup
\PassOptionsToPackage{caption=false}{subfig}
\fi

\usepackage[caption=false]{subfig}

\makeatother

\usepackage{babel}
\begin{document}
\title{Warm Topological Langmuir Cyclotron Wave}
\author{Virginia Billings}
\email{vbilling@ur.rochester.edu }

\affiliation{Department of Physics and Astronomy, University of Rochester, Rochester,
NY 14627}
\author{Hong Qin}
\email{hongqin@princeton.edu }

\affiliation{Department of Astrophysical Sciences, Princeton University, Princeton,
NJ 08540}
\author{Chuang Ren}
\email{chuang.ren@rochester.edu}

\affiliation{Department of Mechanical Engineering and Department of Physics and Astronomy, University of Rochester, Rochester,
NY 14627}

\author{J. B. Marston}
\email{marston@brown.edu }

\affiliation{Brown Center for Theoretical Physics and Innovation and Department of Physics, Brown University, Providence, RI 02912-1843}

\begin{abstract}
Finite-temperature effects in magnetized electron plasmas create a
new Weyl-point degeneracy between the warm Langmuir and right-circularly
polarized waves. The associated topological charge at this warm Weyl
point is found to be 1, which, by the index theorem, predicts a gap-traversing
topological edge mode. Solving the full warm-fluid eigenmode problem
In a 1D inhomogeneous equilibrium, we numerically identify this anticipated
mode as the warm topological Langmuir-cyclotron wave, which is absent
in the cold limit and occurs in a parameter regime relevant to the
LArge Plasma Device (LAPD) at UCLA.
\end{abstract}
\maketitle
The discovery of modes of topological origin has revolutionized condensed matter
physics \citep{Thouless1982,Simon1983,Hasan2010,Qi2011,Armitage2018},
and recent investigations have demonstrated that these concepts are
not limited to quantum systems with crystalline structures but also
applicable to classical continuous media such as neutral fluids \citep{Delplace2017,Perrot2019,Zhu2023,Tong2023,Leclerc2024,Frazier2025a}
and plasmas \citep{Parker2020,Parker2020a,Parker2021,fu2021topological,Qin2023,Fu2022a,Bernety2023,Palmerduca2023,Fu2024,Qin2024Oblique,Rajawat2025,MesaDame2025,MehrpourBernety2024,PalmerducaThesis2025,Xie2025,Rao2025a}.
More broadly, topological
plasma physics is a frontier research area applying the topological
concepts and methods. It has been shown that plasmas can host robust, unidirectional modes
that are topologically protected from scattering and reflection, a
property that is expected to have immense technological potential. A unique property in plasmas is that non-trivial
topology in plasma dynamics manifests in phase space, rather than
the momentum-space Brillouin zone of crystal lattices. Topological
plasma waves in an inhomogeneous system are understood as spectral
flows induced by the nontrivial topology of the bulk eigenmode bundles
over a 2-sphere in phase space surrounding a degeneracy known as a
Weyl point \citep{Marciani2020,Delplace2022,Qin2023,Faure2023,Fu2024,Jezequel2024,Jezequel2025}.
A robust index theorem links the spectral-flow index---the net number
of modes of topological origin---to the topological charge, the Chern number
of the eigenmode bundle, at the phase-space Weyl point \citep{Faure2023}. In physical terms, a bulk spectral degeneracy gives rise to a spatially localized branch in the corresponding inhomogeneous plasma, linking robust edge propagation to bulk wave topology rather than to specific profile or boundary details. 

In cold magnetized plasmas, the background magnetic field breaks time-reversal
symmetry, and two topological modes have been identified: the topological
gaseous plasmon polariton (GPP) \citep{Parker2020} and the topological
Langmuir-cyclotron wave (TLCW) \citep{Qin2023,Fu2022a}. These modes
are tied to the nontrivial topology of eigenmode bundles near the
Weyl points associated with the Langmuir--L and Langmuir--R resonances,
respectively. In the cold magnetized electron plasma system, these
are the only two nontrivial isolated Weyl points.

In this letter, we report a new topological mode in warm magnetized
plasmas: the warm topological Langmuir-cyclotron wave (WTLCW). This
mode is associated with a new Weyl point, a \textquotedblleft warm\textquotedblright{}
Weyl point of degeneracy (resonance) between the Langmuir wave and
the R-wave induced by finite temperature. Using a two-mode projection
method \citep{Qin2023}, we find the topological charge at this warm
Weyl point is $1$, which predicts the existence of a topological edge
mode. The anticipated mode is then identified numerically as the WTLCW
from the full eigenmode system of the warm-fluid equations in an inhomogeneous
equilibrium with two bulk regions, chosen so that the warm Weyl point
lies within the bulk band gap. We will also give the location in the
parameter space where the WTLCW is expected to exist for the LArge
Plasma Device (LAPD) at UCLA.

We adopt the warm-fluid equations of a magnetized electron plasma
in a uniform background field $\mathbf{B}_{0}=B_{0}\hat{\mathbf{z}}$.
Ions provide a stationary neutralizing background, so the dynamical
response is carried by the electron fluid and the electromagnetic
fields. Thermal effects enter through the polytropic law for the pressure.
For bulk modes in a homogeneous equilibrium with constant density
$n_{0}$, effective electron thermal speed $C_{e}$, and polytropic
index $\gamma$, the perturbed pressure is $p_{1}=m\gamma C^{2}_{e}n_{1}$.
It is convenient to introduce the dimensionless warm-compressibility
parameter $\alpha\equiv\sqrt{\gamma}\,C_{e}/c$, which controls the
warm modification of the Langmuir wave and the associated bulk topology.

\begin{figure}[ht]
\centering 
\subfloat[]{\includegraphics[width=5cm]{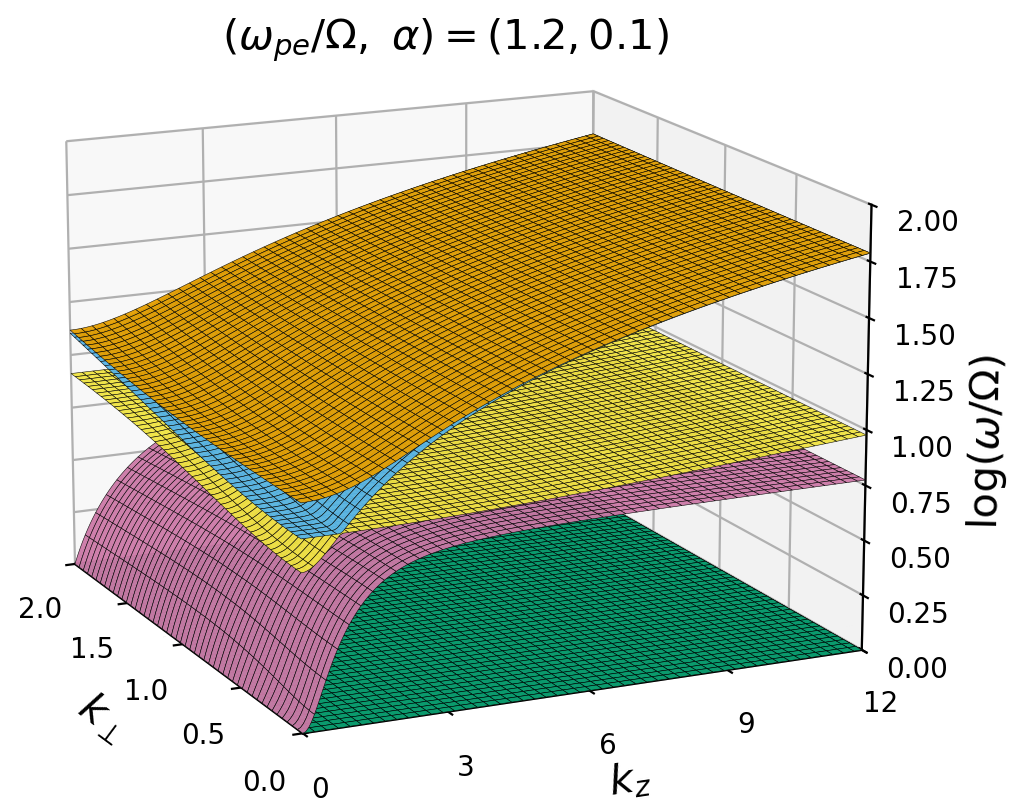}}
\subfloat[]{\includegraphics[width=5cm]{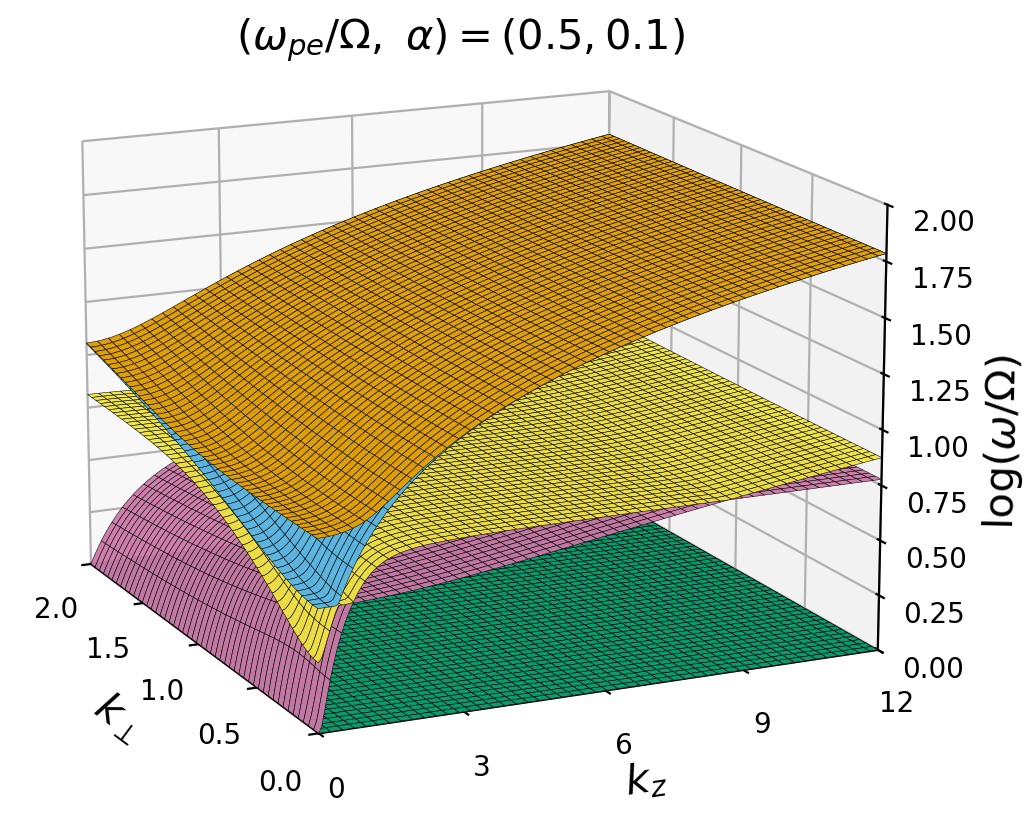}}

\subfloat[]{\includegraphics[width=5cm]{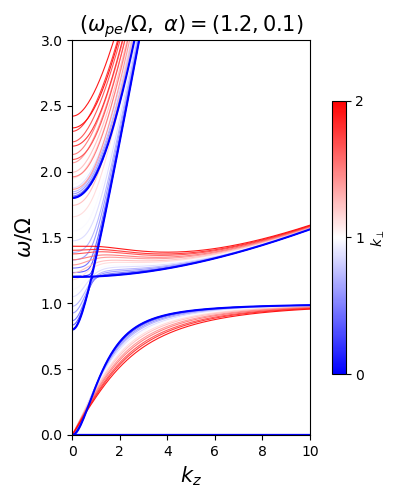}}
\subfloat[]{\includegraphics[width=5cm]{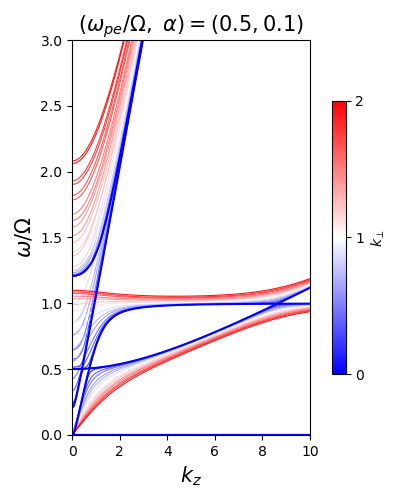}}
\caption{
Spectrum of $H(\mathbf{k})$ defined in Eq. \ref{eqn:H_hermitian}
for a homogeneous warm magnetized plasma. Positive-frequency dispersion
surfaces $\omega_{n}(k_{\perp},k_{z})$ (shown on a log scale) for
representative overdense $(\omega_{p}/\Omega,\alpha)=(1.2,0.1)$ (a)
and underdense $(\omega_{p}/\Omega,\alpha)=(0.5,0.1)$ parameters
(c). Corresponding contour plots (b,d) showing the right- and left-circularly
polarized electromagnetic branches and the warm Langmuir branch; curves
are shown for several $k_{\perp}$ values (color scale). Weyl points
occur in the submanifold of $k_{\perp}=0$ when the Langmuir dispersion
intersects an R/L branch, defining the Weyl points of degeneracy $(\omega^{\ast},k^{\ast}_{z})$.
}
\label{fig:homogweylptmapping}
\end{figure}

Dimensionless units are used in which time is measured in units of
$\Omega^{-1}$ and length in units of $c/\Omega$, with $\Omega\equiv eB_{0}/(mc)$.
The electric and magnetic fields are scaled by a reference amplitude
$\bar{E}$; the velocity perturbation is scaled by $\bar{E}/\sqrt{4\pi mn_{0}},$
and the density perturbation by $(\bar{E}/(c\alpha))\sqrt{n_{0}/(4\pi m)}$.
These scaling symmetrize the coupling between fluid and field variables
and yield a Hermitian linear operator and the standard Hermitian inner
product. For a Fourier component of the linear perturbation relative
to the homogeneous equilibrium at $(\omega,\boldsymbol{k}),$ the
linear system in dimensionless units is

\begin{align}
H(\mathbf{k})|\psi\rangle=\omega|\psi\rangle,\\
H(\mathbf{k})=\begin{bmatrix}0 & \alpha\mathbf{k}\cdot & 0 & 0\\
\alpha\mathbf{k} & i\hat{\mathbf{z}}\times & -i\omega_{p} & 0\\
0 & i\omega_{p} & 0 & -\mathbf{k}\times\\
0 & 0 & \mathbf{k}\times & 0
\end{bmatrix},\label{eqn:H_hermitian}
\end{align}
where $|\psi\rangle=[n_{1},\mathbf{v}_{1},\mathbf{E}_{1},\mathbf{B}_{1}]^{T}$,
$\omega_{p}\equiv\sqrt{4\pi e^{2}n_{0}/m}/\Omega$ is the normalized
plasma frequency (see Appendix \ref{sec:nonuniformoperator}).   The eigensystem is therefore parameterized by $(\omega_{p},\alpha,\mathbf{k})$.
The $10\times10$ Hamiltonian $H(\mathbf{k})$ is Hermitian in the
dimensionless units, and its eigenvalues are real. Diagonalizing $H(\mathbf{k})$
yields ten real dispersion surfaces $\omega_{n}(\mathbf{k})$ over
$\mathbf{k}$-space, shown in Fig.~\ref{fig:homogweylptmapping}.
Two bands lie identically in the $\omega=0$ plane for $\mathbf{k}\neq0$,
representing non-propagating constraint degrees of freedom. Namely,
a longitudinal magnetic component, $\mathbf{B}_{1}\parallel\mathbf{k}$,
associated with the $\nabla\!\cdot\!\mathbf{B}=0$ constraint, and a static electrostatic mode with $\mathbf{E}_{1}\parallel\mathbf{k}$,
or equivalently, $\mathbf{E}_{1}=-i\alpha\,\mathbf{k}\,n_{1}$, are each zero modes. The remaining eight surfaces are dispersive and occur in $\pm\omega$
pairs, with azimuthal symmetry about the $\mathbf{B}_{0}$ axis (i.e.\ dependence
on $k_{\perp}$ and $k_{z}$, where $k_{\perp}\equiv\sqrt{k^{2}_{x}+k^{2}_{y}}$). These bands are the warm-plasma electromagnetic spectrum.

As seen in Fig.~\ref{fig:homogweylptmapping}, isolated Weyl points
occur only in the submanifold of $k_{\perp}=0$, where the transverse
spectrum reduces to the right- and left-circularly polarized ($R/L$)
waves, while the longitudinal sector reduces to the warm Langmuir
wave. It can be rigorously established that Weyl points of the system
only exist in the $k_{\perp}=0$ submanifold, and they are isolated.
In unnormalized form, spectrum in the $k_{\perp}=0$ submanifold are
defined by

\begin{align}
\frac{c^{2}k^{2}_{z}}{\omega^{2}} & =1-\frac{\omega^{2}_{p}}{\omega^{2}-\Omega^{2}}\mp\frac{\Omega}{\omega}\frac{\omega^{2}_{p}}{\omega^{2}-\Omega^{2}},\label{eqn:RLandparallel}\\
\omega^{2} & =\omega^{2}_{p}+\gamma C^{2}_{e}k^{2}_{z},\label{eqn:warmLangmuir}
\end{align}
where the upper/lower sign in Eq.\,(\ref{eqn:RLandparallel}) corresponds
to the $R/L$ branch, respectively, and Eq.\,(\ref{eqn:warmLangmuir})
is the dispersion relation of the warm Langmuir wave. Degeneracies
occur when the warm Langmuir curve intersects either electromagnetic
branch at $(\omega^{*},k^{*}_{z})$. In the underdense regime ($\omega_{p}/\Omega<1$),
three distinct Weyl points are presented, ordered as $k^{*(u)}_{z,L}<k^{*(u)}_{z,R}<k^{*(u)}_{z,R+}$.
In the overdense regime ($\omega_{p}/\Omega>1$), there is only one
Weyl point at $k^{*(o)}_{z,L}$. In the cold limit ($\alpha\to0$),
Eq.\,(\ref{eqn:warmLangmuir}) reduces to the usual plasma frequency,
and $k^{*(u)}_{z,L}$, $k^{*(u)}_{z,R}$, and $k^{*(o)}_{z,L}$ reduce
to their cold counterparts in Refs.~\citep{Parker2020,fu2021topological}.
The warm term $\gamma C^{2}_{e}k^{2}_{z}$ shifts the values of $k^{*(u)}_{z,L}$,
$k^{*(u)}_{z,R}$, and $k^{*(o)}_{z,L}$ relative to the cold case
without changing their characteristics. What is qualitatively different
is that the warm term introduces a new Weyl point $k^{*(u)}_{z,R+}$
at a higher value in the underdense regime. We note that finite temperature
must introduce the second resonance between the warm Langmuir wave
and the R-wave because the latter starts from $0$ at $k_{\perp}=0$,
and is asymptotic to the electron gyrofrequency at large $k_{\perp}$.
But the former stems from plasma frequency and is unbounded for large
$k_{z}$ in the underdense regime. The warm Weyl point $k^{*(u)}_{z,R+}$
does not have a cold counterpart. 

In the cold limit, the Weyl points $k^{*(u)}_{z,L}$and $k^{*(u)}_{z,R}$
are associated with the GPP \citep{Parker2020} and TLCW \citep{Qin2023},
respectively. Is there a topological mode associated with the warm
Weyl point $k^{*(u)}_{z,R+}$? The answer is affirmative, and the
topological mode is WTLCW. To show this, we first calculated the Chern
numbers of the eigenmode bundles on an $\mathbb{S}^{2}$ surface in
the $(k_{z},k_{y},\omega_{p})$ space enclosing the warm Weyl
point $(k_{z,}k_{y},\omega_{p})=(k^{*(u)}_{z,R+},0,\omega^{*})$,
which is known as the topological charge and linked to existence of
the WTLCW in an inhomogeneous system via the index theorem \citep{Faure2023,Qin2023}.
Following Ref.~\citep{Qin2023}, we project the system onto the two-mode
subspace at the warm Weyl point and evaluate the Chern number on
a small enclosing surface. At the warm Weyl point, the system is characterized
by a two-fold degeneracy,

\begin{align}
H(k^{*(u)}_{z,R+},0,\omega^{*})|\psi_{a}\rangle=\omega^{*}|\psi_{a}\rangle,\qquad H(k^{*(u)}_{z,R+},0,\omega^{*})|\psi_{b}\rangle=\omega^{*}|\psi_{b}\rangle,
\end{align}
with $\{|\psi_{a}\rangle,|\psi_{b}\rangle\}$ spanning the degenerate
eigenspace. For $(k_{z,}k_{y},\omega_{p})$ on a small sphere
$\mathbb{S}^{2}$ enclosing $(k^{*(u)}_{z,R+},0,\omega^{*})$, the
dynamics are dominated by the projected $2\times2$ matrix

\begin{align}
W_{ij}(k_{z,}k_{y},\omega_{p})=\langle\psi_{i}|\,H(k_{z,}k_{y},\omega_{p})|\psi_{j}\rangle,\qquad i,j\in\{a,b\}.\label{eqn:two_mode_proj}
\end{align}
Diagonalizing $W(k_{z,}k_{y},\omega_{p})$ yields two projected
eigenvectors $|\psi^{\pm}(k_{z,}k_{y},\omega_{p})\rangle$ defined
on $\mathbb{S}^{2}$. The projected eigenvalues remain nondegenerate
on $\mathbb{S}^{2}$ because the Weyl point is isolated. Each band
carries an integer Chern number given by the integral

\begin{align}
C_{\pm} & =\frac{1}{2\pi}\int_{\mathbb{S}^{2}}\left(\nabla_{\mathbf{k}}\times\mathbf{A}_{\pm}\right)\cdot d\mathbf{S}\label{eqn:chern_proj}
\end{align}
where $\mathbf{A}_{\pm}(k_{z,}k_{y},\omega_{p})=i\langle\psi^{\pm}|\nabla\psi^{\pm}\rangle$
is the Berry connection. Calculation (see Appendix \ref{sec:numweylcharge})
shows that $C_{\pm}=\mp1$ at the warm Weyl point $(k_{z,}k_{y},\omega_{p})=(k^{*(u)}_{z,R+},0,\omega^{*})$
. We briefly note that the Chern numbers at all other Weyl points
are unchanged from their cold-plasma counterparts, indicating that
the GPP and TCLW remain topologically identical in warm and cold plasmas.

According to the index theory, the non-vanishing Chern number $C_{\pm}$
of the eigenmode bundles over a small sphere $\mathbb{S}^{2}$ enclosing
the warm Weyl point in the parameter space $(k_{z,}k_{y},\omega_{p})$
is equal to the spectral flow index in an inhomogeneous equilibrium
consisting of two bulk regions when the warm Weyl point lies in the band
gap of the bulk regions \citep{Faure2023,Qin2023}. The spectral flow
is the anticipated WTLCW. 

We now identify the mode by numerically solving the eigenvalue problem
in such a 1D inhomogeneous equilibrium with density and temperature
profiles $n_{0}(x)$ and $T_{0}(x)$ chosen so that $n_{0}(x)T_{0}(x)=\mathrm{const}$.
The equilibrium density is taken as

\begin{align}
n_{0}(x)=\frac{(n_{h}-n_{l})}{2}\left[\tanh{\left(-\frac{x-l}{\sigma}\right)}+\tanh{\left(\frac{x+l}{\sigma}\right)}\right]+n_{l},\label{eqn:densityprofile}
\end{align}
with center bulk values $n_{h}$ and edge bulk values $n_{l}$ connected
by transition layers centered at $x=\pm l$ of thickness $\sigma$.
The local plasma frequency $\omega_{p}(x)$ and warm parameter $\alpha(x)$
then vary across the slab through their dependence on $n_{0}(x)$.
At fixed $k_{z}$, the locations of the warm Weyl points are given
by the roots of

\begin{equation}
r(x)\equiv\omega_{p}(x)-\omega^{\ast}_{p}\!\big(k_{z};C_{e}(x)\big),
\end{equation}
where $\omega^{\ast}_{p}(k_{z};C_{e})$ is the plasma frequency at
which the warm Langmuir wave resonates with the R-wave for the local
thermal parameter $C_{e}$ according to Eqs.\,(\ref{eqn:RLandparallel})
and (\ref{eqn:warmLangmuir}). The equilibrium profile is constructed
such that the roots of $r(x)$ situate in the transition region near
$x=\pm l.$ Linearization of the warm-fluid and Maxwell equations
about this inhomogeneous equilibrium gives
\begin{align}
i\partial_{t}\psi=\hat{H}\psi,\\
\hat{H}(\mathbf{r},-i\eta\nabla)=\begin{bmatrix}0 & A & 0 & 0\\
B & i\hat{\mathbf{z}}\times & -i\omega_{p}(x) & 0\\
0 & i\omega_{p}(x) & 0 & i\eta\nabla\times\\
0 & 0 & -i\eta\nabla\times & 0
\end{bmatrix},\label{eqn:Hhat}
\end{align}
where the density--velocity coupling operators are
\begin{align}
A & =-\alpha(x)\,(i\eta\nabla)\cdot+(i\eta\nabla\alpha(x))\cdot,\label{eqn:A_Hhat}\\
B & =-\alpha(x)\,(i\eta\nabla)-(i\eta\nabla\alpha(x)).\label{eqn:B_Hhat}
\end{align}

For a detailed derivation of Eqs.~\eqref{eqn:Hhat}-\eqref{eqn:B_Hhat}, see Appendix \ref{sec:nonuniformoperator}. Relative to the homogeneous equations, the new structure is the gradient
terms $\nabla\alpha(x)$, which enter at the same order in $\eta$
as the differential operators,  and the Hamiltonian $\hat{H}(\mathbf{r},-i\eta\nabla)$
is Hermitian. Because the equilibrium varies only in $x$, the
Fourier components in $y$ and $z$ are decoupled, and we let $\psi(\mathbf{r})=\psi(x)\exp[i(k_{y}y+k_{z}z)]$
so that $-i\eta\partial_{y}\to k_{y}$ and $-i\eta\partial_{z}\to k_{z}$,
while $-i\eta\partial_{x}$ remains an operator. Under these substitutions,
$\hat{H}(\mathbf{r},-i\eta\nabla)$ reduces to the one-dimensional
partial-differential operator $\hat{H}(x,-i\eta\partial_{x},k_{y},k_{z})$,
which defines the slab eigenvalue problem at fixed wavenumbers $(k_{y},k_{z})$,

\begin{equation}
\hat{H}(x,-i\eta\partial_{x},k_{y},k_{z})\,\psi(x)=\omega\,\psi(x).
\end{equation}
The spectrum of $\hat{H}(x,-i\eta\partial_{x},k_{y},k_{z})$ gives
the global modes of the inhomogeneous plasma. 

In theoretical analysis, especially the proof of the index theorem,
scale separation between equilibrium gradients and wave structure
is often assumed. Let $L\sim\left|n_{0}/\partial_{x}n_{0}\right|$
denote the characteristic transverse inhomogeneity length. Scale separation
implies that $\eta\equiv c/(L\Omega)\ll1$. In the Weyl-quantized
(pseudo-differential) formulation, $\eta$ plays the role of a semiclassical
parameter governing how spatial variation enters the operator $\hat{H}(\mathbf{r},-i\eta\nabla)$.
For numerical solutions, it is not necessary to assume scale separation.
It is expected that the numerical results agree with the analytical
findings when the semiclassical condition is satisfied. 

\begin{figure}[ht]
\subfloat[]{\includegraphics[width=5cm]{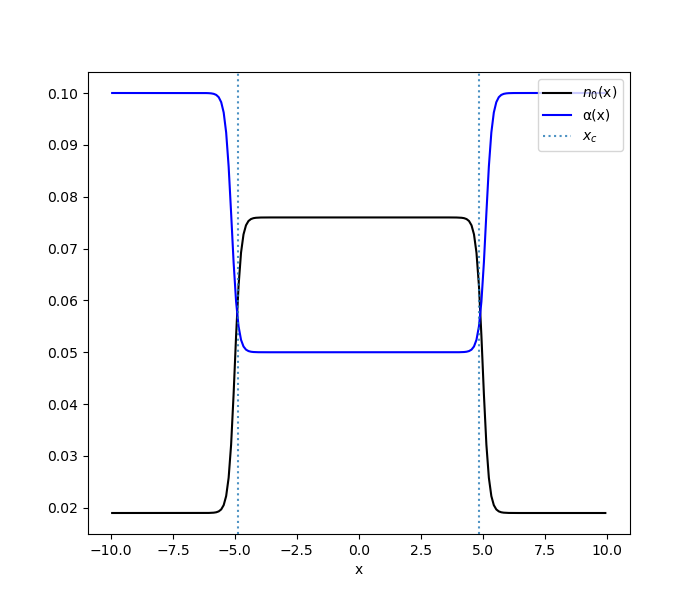}

}\subfloat[]{\includegraphics[width=5cm]{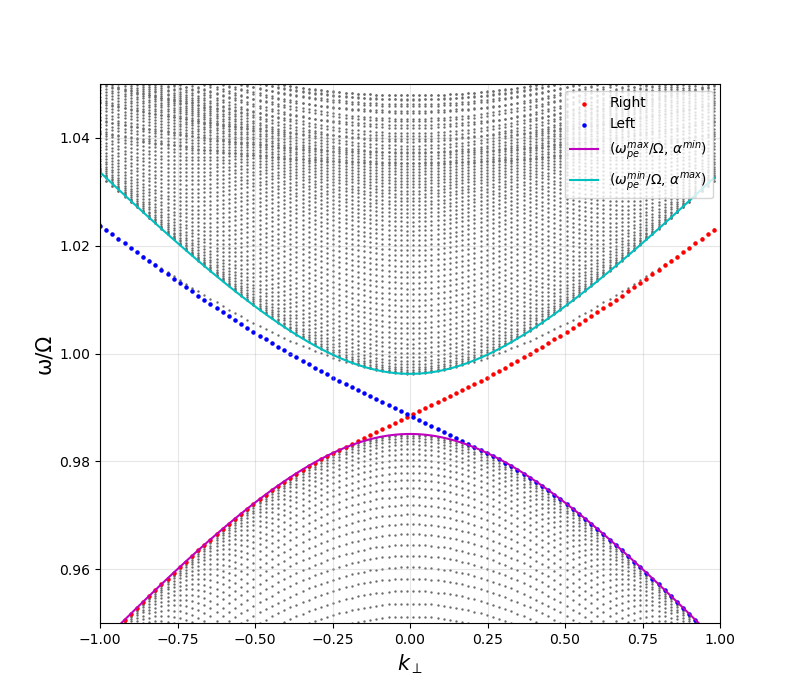}

}

\subfloat[]{\includegraphics[width=5cm]{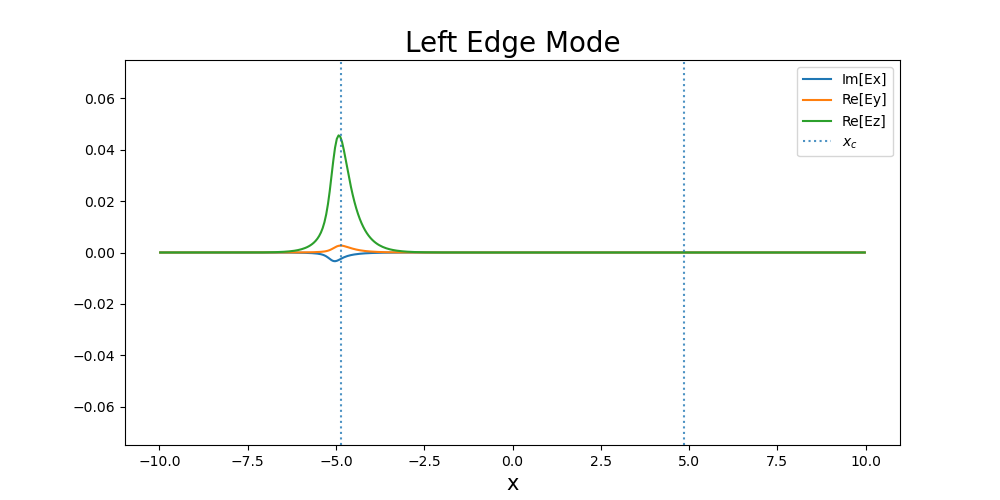}

}\subfloat[]{\includegraphics[width=5cm]{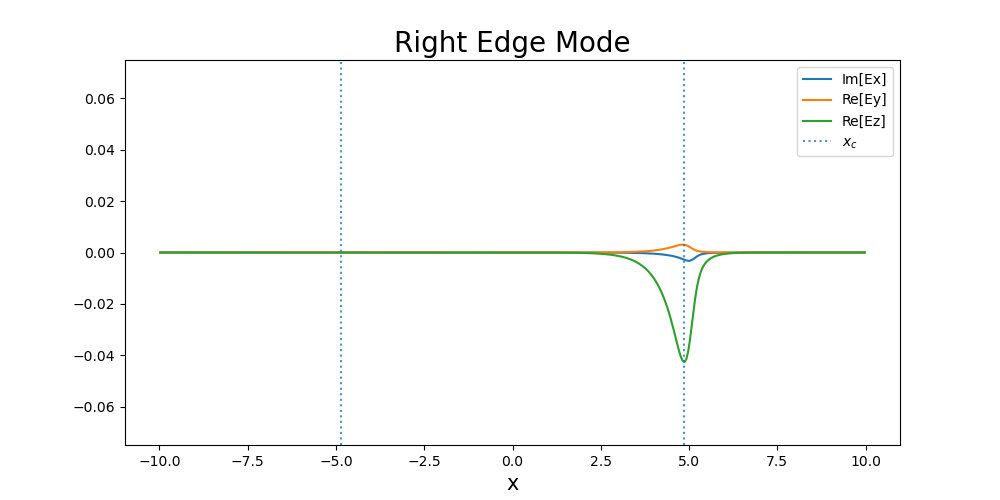}

}

\caption{Emergence of the WTLCM at $k_{z}=8.0$ with finite pressure $p_{0}=1.9\times10^{-4}$.
Profiles $\omega_{p}(x)$ and $\alpha(x)$, together with the required
crossing frequency $\omega^{\ast}_{p}(k_{z};C_{e}(x))$ from the warm
on-axis crossing (a); intersections $r(x)\equiv\omega_{p}(x)-\omega^{\ast}_{p}(k_{z};C_{e}(x))=0$
determine the predicted interfaces $x^{\pm}_{c}$ (dashed lines).
The spectrum consists of three parts: the continuous region (black)
and two spectral flows (red and blue) of the WTLCW (b). The edge-localized
eigenmode structures of the WTLCW at the left (c) and right (d) interface
regions. }

\label{fig:kz8thermal}
\end{figure}

Figure~\ref{fig:kz8thermal} shows the numerically calculated spectrum
for a 1D equilibrium (a) specified by Eq.\,(\ref{eqn:densityprofile})
with $n_{h}= 0.076$, $n_{l}=0.019$, and $p_{0}=1.9\times10^{-4}$. The spectrum
at $k_{z}=8.0$ is plotted against $k_{y}$ in (b), which consists
of three parts. The spectrum of the inhomogeneous system plotted as
black dots is the continuous spectrum corresponding to the bulk modes
of the high and low density regions. The red and blue curves traversing
the bulk gap are the spectral flows of the WTLCW. Their eigenfunctions
are plotted in (c) and (d), showing edge-localized structure centered
at the predicted locations $x^{\pm}_{c}$ (dashed lines). The blue/red
curve is the anticipated WTLCW at the left/right interface region.
The fact that there is one WTLCW at each interface agrees with the
result of $C_{\pm}=\mp1$ at the warm Weyl point, as required by the
index theorem \citep{Faure2023} or bulk-edge correspondence. 

\begin{figure}[t]
\subfloat[]{\includegraphics[width=5cm]{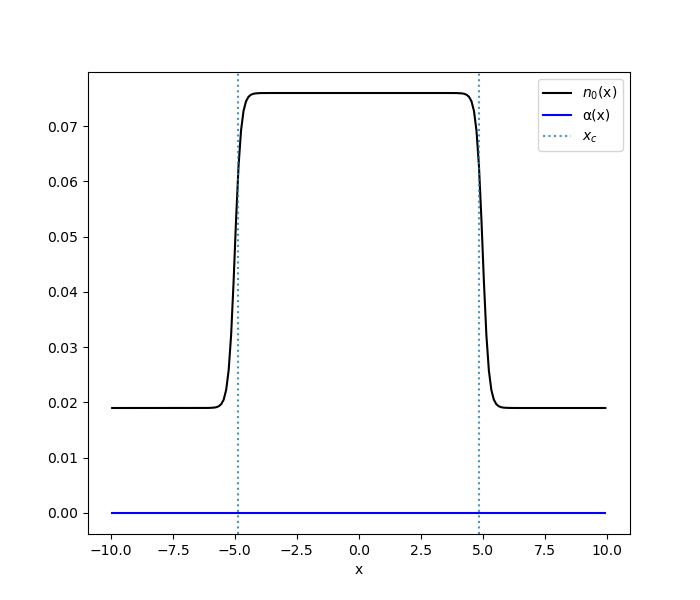}

}\subfloat[]{\includegraphics[width=5cm]{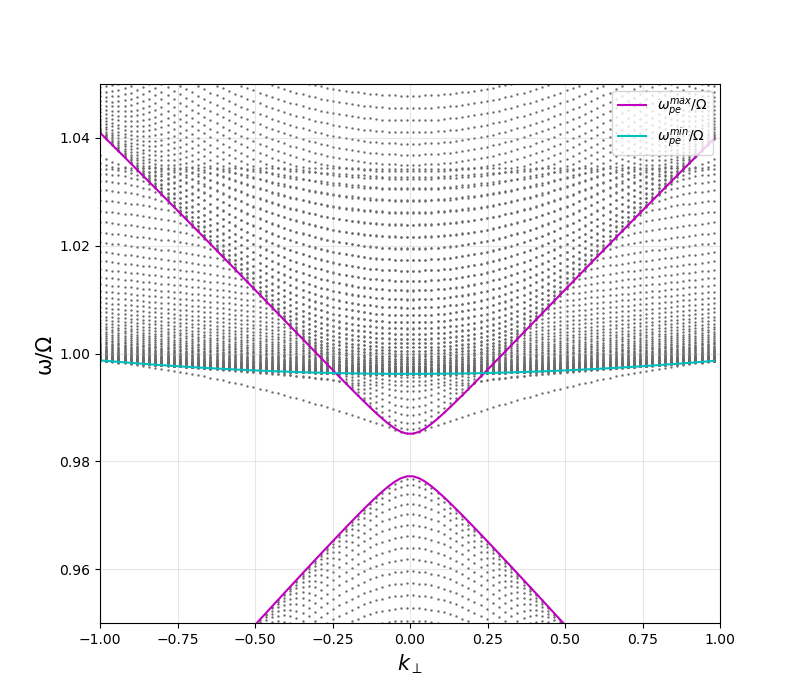}

}\caption{Cold-limit null test at $k_{z}=8.0$ and $p_{0}=0$ ($\alpha\equiv0$)
with the same density profile as Fig.~4. (a) The warm Weyl condition
has no root within the slab for the chosen density range (the required
$\omega^{\ast}_{p}$ lies above the maximum $\omega_{p}$ attainable).
(b) Corresponding slab spectrum lacks a gap-traversing spectral flow.}
\label{fig:kz8nonthermal}
\end{figure}

As a null test, we set $p_{0}=0$ ($\alpha\equiv0$) at the same $k_{z}$
and density profile as shown in Fig.\,\ref{fig:kz8nonthermal}(a).
In this limit, the warm Weyl condition has no root within the slab,
and no corresponding interface branch appears in the spectrum shown
in Fig.~\ref{fig:kz8nonthermal}(b), demonstrating that the WTLCW
is created by the finite-temperature induced Weyl point and vanishes
in the cold limit.

Because WTLCW is a topological mode, it enjoys the robustness associated
with topological protection. It's expected to propagate in a unidirectional
manner without scattering and reflection along the interface defined
by the warm Weyl point, as in the cases of TLCW \citep{Qin2023} and
GPP \citep{Parker2020}.  This desirable property can be explored as
an effective mechanism for heating and particle acceleration. 

\begin{figure}[ht]
\subfloat[]{\includegraphics[width=5cm]{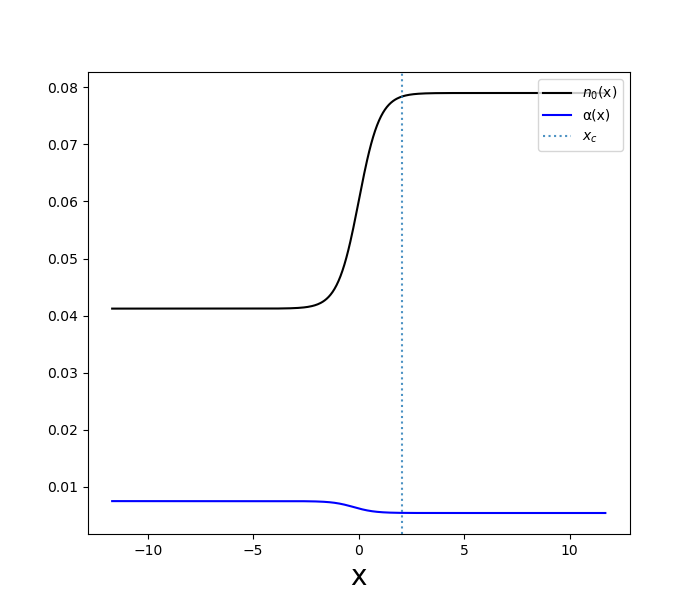}

}\subfloat[]{\includegraphics[width=5cm]{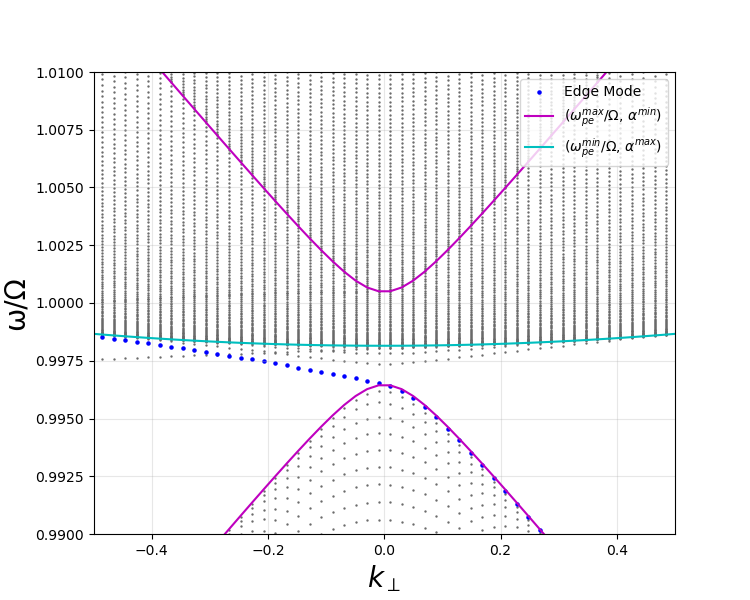}

}\subfloat[]{\includegraphics[width=5cm]{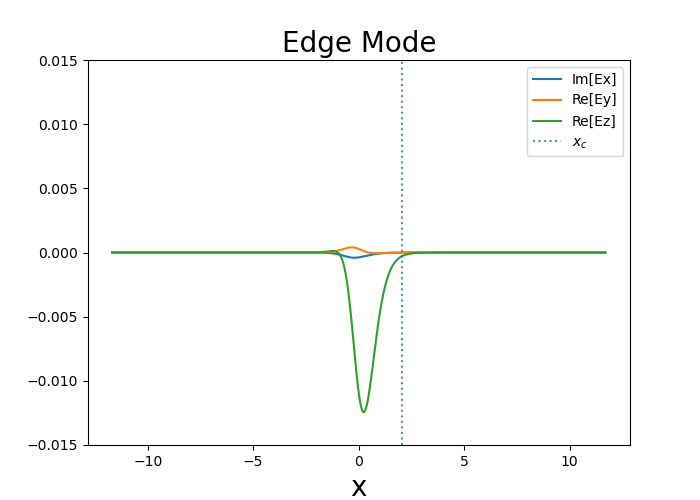}

}\caption{Predicated WTLCM for machine parameters typical to the LAPD. The inhomogeneous
equilibrium is made of two bulks and an interface region (a). The
continuous spectrum (black) corresponds to the bulk modes of the homogeneous
bulks and the spectral flow (red) is the WTLCW (b). The mode structure
of the WTLCW is localized in the interface region. }

\label{fig:LAPD}
\end{figure}

Next, we give a numerically calculated WTLCW using the machine parameters
typical to the LAPD at UCLA. The typical density,
magnetic field, and temperature of the LAPD plasma are $n_{h}=0.385413\times10^{12}\,\mathrm{cm^{-3}}$,
$B_{0}=2\times10^{3}\,\mathrm{G}$, and $T_{0}=15~\mathrm{eV}$. We
assume that in the edge region, the plasma density is dropped to $n_{l}=0.2\times10^{12}\,\mathrm{cm^{-3}}$.
To maintain the equilibrium, the temperature is increased towards
the edge accordingly. Because this system is in a cylindrical geometry,
there is only one density transition region. The equilibrium model
is plotted in Fig.\,\ref{fig:LAPD}(a). For the case of $k_z=19.7 \ \text{cm}^{-1}$, the numerically solved spectrum
of the system is shown in Fig.\,\ref{fig:LAPD}(b). Akin to Fig.\,\ref{fig:kz8thermal},
the spectrum consists of a continuum (black) corresponding to the
bulk modes of the homogeneous regions and the spectral flow (red)
of the WTLCW. The eigenmode structure of the WTLCW shown in Fig.\,\ref{fig:LAPD}(c)
is localized in the interface region, as expected. 

In conclusion, finite temperature does more than perturb the cold-plasma
spectrum: it creates a genuinely new Weyl-point degeneracy in a warm
magnetized electron plasma, and thereby enables a new topological
edge mode, the warm topological Langmuir-cyclotron wave (WTLCW). By
projecting the warm-fluid eigenmode system onto the two-mode degenerate
subspace at the warm Weyl point, our calculation shows that its topological
charge is $1$, which, via bulk-edge correspondence, predicts a gap-crossing
spectral flow in an inhomogeneous equilibrium. Full eigenmode calculations
in an inhomogeneous equilibrium with two bulk regions confirm this
prediction by producing an interface-localized mode that traverses
the bulk gap, while an otherwise identical cold-limit calculation
provides a clean null test with no such mode. These results establish
a temperature-enabled route to topological waves in magnetized plasmas
and motivate experimental searches in accessible parameter regimes,
with potential applications to robust particle energization mechanism
by topological waves along engineered density interfaces.
\begin{acknowledgments}
VB and CR were partially supported by DARPA under Award No. N66001242007. JBM is partially supported by a grant from the Simons Foundation International [SFI-PD-Pivot Mentor-00008573, JBM]. Additional gratitude is extended to Dr.\ Chaitanya Murthy, Dr.\ Sree Harsha Naropanth Ramamurthy, Elisha Haber, and Luke Anderson for valuable discussions.
\end{acknowledgments}

\appendix
\section{Derivation of the nonuniform warm-fluid operator}
\label{sec:nonuniformoperator}

We consider an electron plasma with immobile neutralizing ions in a uniform
background magnetic field, $\mathbf B_0 = B_0 \hat{\mathbf z}.$ The equilibrium density and temperature vary in the transverse direction, $n_0 = n_0(x), \ T_0 = T_0(x),$ subject to the equilibrium condition $n_0(x)T_0(x)=\mathrm{const}.$ Thus the equilibrium pressure is spatially uniform, $p_0 \propto n_0(x)T_0(x)=\mathrm{const}, \ \nabla p_0 = 0.$ We take 
\begin{align}
    n(\mathbf r,t) &= n_0(x)+n_1(\mathbf r,t),\\
    p(\mathbf r,t) &= p_0+p_1(\mathbf r,t),\\
    \mathbf v(\mathbf r,t) &= \mathbf v_1(\mathbf r,t),\\
    \mathbf E(\mathbf r,t) &= \mathbf E_1(\mathbf r,t),\\
    \mathbf B(\mathbf r,t) &= B_0\hat{\mathbf z}+\mathbf B_1(\mathbf r,t).
\end{align}
Linearizing the warm-fluid-Maxwell system about this equilibrium gives
\begin{align}
    \partial_t n_1
    &=
    -\nabla\cdot\left[n_0(x)\mathbf v_1\right],
    \label{eq:nonuniform_continuity_dimensional}
    \\
    \partial_t \mathbf v_1
    &=
    -\frac{e}{m}\mathbf E_1
    +\Omega\,\hat{\mathbf z}\times \mathbf v_1
    -\frac{1}{m n_0(x)}\nabla p_1,
    \label{eq:nonuniform_momentum_dimensional}
    \\
    \partial_t \mathbf B_1
    &=
    -c\nabla\times \mathbf E_1,
    \label{eq:nonuniform_faraday_dimensional}
    \\
    \partial_t \mathbf E_1
    &=
    c\nabla\times \mathbf B_1
    +\frac{m}{e}\omega_p^2(x)\mathbf v_1,
    \label{eq:nonuniform_ampere_dimensional}
\end{align}
where
\begin{align}
    \Omega = \frac{eB_0}{mc},
    \qquad
    \omega_p(x)=\sqrt{\frac{4\pi e^2 n_0(x)}{m}}.
\end{align}

For the pressure perturbation, we close the moment hierarchy with a local
adiabatic warm-fluid closure,
\begin{align}
    p_1
    =
    m\gamma C_e^2(x)n_1
    =
    \gamma \frac{p_0}{n_0(x)} n_1.
\end{align}
Here, \(C_e^2(x)=p_0/[m n_0(x)]\), where \(C_e(x)\) is the local equilibrium thermal speed.
Thus, the pressure response is locally proportional to the density perturbation,
while the equilibrium pressure itself remains spatially uniform. It is useful to introduce the local thermal parameter
\begin{align}
    \alpha(x)
    =
    \sqrt{\frac{\gamma C_e^2(x)}{c^2}}
    =
    \sqrt{\frac{\gamma p_0}{m c^2 n_0(x)}},
    \label{eq:alpha_def_nonuniform}
\end{align}
so that $p_1=m c^2 \alpha^2(x)n_1.$
We now nondimensionalize using the electron cyclotron frequency and the
equilibrium inhomogeneity scale \(L\),
\begin{align}
    t'=\Omega t,
    \qquad
    \mathbf r'=\frac{\mathbf r}{L},
    \qquad
    \nabla' = L\nabla,
    \qquad
    \eta=\frac{c}{L\Omega}.
\end{align}
The parameter \(\eta\) measures the ratio of the electromagnetic length scale
\(c/\Omega\) to the equilibrium scale length \(L\) \cite{Qin2023}. The fields and fluid
variables are scaled as
\begin{align}
    \mathbf E_1'
    =
    \frac{\mathbf E_1}{\bar E},
    \qquad
    \mathbf B_1'
    =
    \frac{\mathbf B_1}{\bar E},
    \qquad
    \mathbf v_1'
    =
    \frac{\sqrt{4\pi m n_0(x)}}{\bar E}\,\mathbf v_1,
    \qquad
    n_1'
    =
    \frac{c\alpha(x)}{\bar E}
    \sqrt{\frac{4\pi m}{n_0(x)}}\,n_1.
\end{align}

Because the fluid variables are normalized using the local equilibrium
profiles, spatial derivatives generate profile-gradient corrections. Written
in terms of \(\alpha(x)\), these corrections give the derivative terms shown below. After dropping primes, the normalized linearized system can be written as

\begin{align}
    i\partial_t n_1
    &=
    -\alpha(x)(i\eta\nabla)\cdot\mathbf v_1
    +(i\eta\nabla\alpha)\cdot\mathbf v_1,
    \label{eq:nonuniform_continuity_normalized}
    \\
    i\partial_t \mathbf v_1
    &=
    -i\omega_p(x)\mathbf E_1
    +i\hat{\mathbf z}\times\mathbf v_1
    -\alpha(x)(i\eta\nabla)n_1
    -(i\eta\nabla\alpha)n_1,
    \label{eq:nonuniform_momentum_normalized}
    \\
    i\partial_t \mathbf B_1
    &=
    -i\eta\nabla\times\mathbf E_1,
    \label{eq:nonuniform_faraday_normalized}
    \\
    i\partial_t \mathbf E_1
    &=
    i\eta\nabla\times\mathbf B_1
    +i\omega_p(x)\mathbf v_1.
    \label{eq:nonuniform_ampere_normalized}
\end{align}

Defining the state vector $\psi =[n_1, \mathbf v_1, \mathbf E_1, \mathbf B_1]^{T},$ the system takes the form $i\partial_t\psi = \hat H(\mathbf r,-i\eta\nabla)\psi,$ with
\begin{align}
\hat{H}(\mathbf{r},-i\eta\nabla)=\begin{bmatrix}0 & A & 0 & 0\\
B & i\hat{\mathbf{z}}\times & -i\omega_{p}(x) & 0\\
0 & i\omega_{p}(x) & 0 & i\eta\nabla\times\\
0 & 0 & -i\eta\nabla\times & 0
\end{bmatrix},\label{eq:Hhat_nonuniforma}
\end{align}
where the density--velocity coupling operators are
\begin{align}
A & =-\alpha(x)\,(i\eta\nabla)\cdot+(i\eta\nabla\alpha(x))\cdot,\label{eqn:A_Hhata}\\
B & =-\alpha(x)\,(i\eta\nabla)-(i\eta\nabla\alpha(x)).\label{eqn:B_Hhata}
\end{align}

Following the Weyl-quantization framework of Ref.~\cite{Qin2023}, we regard \(\hat H(\mathbf r,-i\eta\nabla)\) as a pseudo-differential operator obtained
from a bulk Hamiltonian symbol \(H(\mathbf r,\mathbf k)\),
\begin{align}
    \hat H(\mathbf r,-i\eta\nabla)
    =
    \mathrm{Op}_{\eta}\!\left[H(\mathbf r,\mathbf k)\right],
    \qquad
    \hat{\mathbf k}=-i\eta\nabla .
\end{align}
The corresponding local bulk Hamiltonian is therefore obtained by replacing
the differential operator by its Weyl symbol,
\(-i\eta\nabla\rightarrow \mathbf k\), giving \(H(x,\mathbf k)\). For the inhomogeneous slab geometry used to identify interface modes,
translational symmetry in \(y\) and \(z\) is retained while the \(x\)-dependence
is kept explicitly. Thus \(k_y\) and \(k_z\) are treated as parameters, while
\(k_x\) is quantized according to \(k_x\rightarrow -i\eta\partial_x\), yielding
\(\hat H(x,-i\eta\partial_x,k_y,k_z)\). In the homogeneous limit,
where \(\nabla\alpha\rightarrow 0\) and
\(\omega_p(x)\rightarrow \omega_p\), the equilibrium parameters become
spatially uniform and
Eqs.~\eqref{eq:nonuniform_continuity_normalized}--\eqref{eq:nonuniform_ampere_normalized}
reduce to the homogeneous warm-fluid system,

\begin{align}
    \omega n_1 &= \alpha \mathbf k \cdot \mathbf v_1, \\
    \omega \mathbf v_1
    &= -i\omega_p \mathbf E_1
    + i\hat{\mathbf z}\times \mathbf v_1
    + \alpha n_1 \mathbf k, \\
    \omega \mathbf B_1 &= \mathbf k \times \mathbf E_1, \\
    \omega \mathbf E_1
    &= -\mathbf k \times \mathbf B_1
    + i\omega_p \mathbf v_1.
\end{align}

Consequently, Eq.~\eqref{eq:Hhat_nonuniforma} reduces to the homogeneous
warm-fluid Hamiltonian,

\begin{align}
    H(\mathbf k) =
    \begin{bmatrix}
        0 & \alpha \mathbf k \cdot & 0 & 0 \\
        \alpha \mathbf k & i\hat{\mathbf z}\times & -i\omega_p & 0 \\
        0 & i\omega_p & 0 & -\mathbf k\times \\
        0 & 0 & \mathbf k\times & 0
    \end{bmatrix}.
    \label{eqn:homogenousH}
\end{align}

This Hamiltonian satisfies the Hermitian eigenvalue problem, $H(\mathbf k)\psi = \omega \psi$, with state vector, $\psi = [\, n_1,\ \mathbf v_1,\ \mathbf E_1,\ \mathbf B_1 \,]^T.$ Equation \eqref{eqn:homogenousH} is consistent, up to normalization and ordering conventions, with the finite-temperature electron magnetofluid operator used in Ref.~\cite{Rao2025a}.

\section{Numerical evaluation of the Weyl charge}
\label{sec:numweylcharge}

We compute the Chern numbers associated with an isolated two-mode degeneracy using the gauge-invariant lattice formulation of Fukui, Hatsugai, and Suzuki (FHS)~\cite{Fukui2005}. For a normalized nondegenerate eigenstate $|\psi(\boldsymbol q)\rangle$ defined on a two-dimensional parameter surface, the continuum Berry connection is
\begin{align}
    A_\mu(\boldsymbol q)
    =
    i\langle \psi(\boldsymbol q)|\partial_\mu \psi(\boldsymbol q)\rangle ,
\end{align}
where $\mu=1,2$ labels local coordinates on the surface. On a discrete mesh, the corresponding $U(1)$ link variable is defined by
\begin{align}
    U_\mu(p)
    =
    \frac{\langle \psi(p)|\psi(p+\hat\mu)\rangle}
    {|\langle \psi(p)|\psi(p+\hat\mu)\rangle|}.
\end{align}
This expression follows from expanding the neighboring state,
\begin{align}
    |\psi(p+\hat\mu)\rangle
    =
    |\psi(p)\rangle
    +
    \Delta_\mu \partial_\mu |\psi(p)\rangle
    +
    \mathcal O(\Delta_\mu^2),
\end{align}
so that the overlap contains the Berry phase accumulated along the mesh edge. Under the gauge transformation
\begin{align}
    |\psi(p)\rangle \rightarrow e^{i\chi(p)}|\psi(p)\rangle ,
\end{align}
the link transforms as
\begin{align}
    U_\mu(p)
    \rightarrow
    e^{-i\chi(p)}
    U_\mu(p)
    e^{i\chi(p+\hat\mu)} ,
\end{align}
while products around closed plaquettes are gauge-invariant.

For an isolated on-axis degeneracy at $\mathbf k^\ast=(0,0,k_z^\ast),$
we choose a small closed surface in local Cartesian coordinates $\delta\mathbf k=(\delta k_x,\delta k_y,\delta k_z)$
centered at $\mathbf k^\ast$. In the numerical calculation, this surface is the boundary of a cube, $\partial[-\epsilon,\epsilon]^3,$
with outward-oriented faces. The cube size $\epsilon$ is chosen so that the targeted degeneracy is enclosed and the projected two-mode gap remains open everywhere on the boundary.

At each point on the enclosing surface, the full Hermitian matrix is projected onto the two-dimensional degenerate subspace at $\mathbf k^\ast$,
\begin{align}
    W(\delta\mathbf k)
    =
    \Omega^\dagger H(\mathbf k^\ast+\delta\mathbf k)\Omega ,
\end{align}
where the columns of $\Omega$ form an orthonormal basis for the degenerate subspace. Diagonalizing $W(\delta\mathbf k)$ gives the upper and lower projected eigenvectors,
$|\psi_+(\delta\mathbf k)\rangle$ and $|\psi_-(\delta\mathbf k)\rangle$.

The FHS construction is then applied separately to each branch. For an oriented plaquette based at $p$, the gauge-invariant plaquette product is
\begin{align}
    \mathcal W_\pm(p)
    =
    U_{1,\pm}(p)
    U_{2,\pm}(p+\hat 1)
    U^{-1}_{1,\pm}(p+\hat 2)
    U^{-1}_{2,\pm}(p),
\end{align}
and the discrete Berry flux is
\begin{align}
    F_\pm(p)
    =
    \arg \mathcal W_\pm(p),
\end{align}
with $\arg$ evaluated on the principal branch $(-\pi,\pi]$. The Chern number is obtained by summing over all plaquettes on the six outward-oriented faces,
\begin{align}
    C_\pm
    =
    \frac{1}{2\pi}
    \sum_{p\in \partial[-\epsilon,\epsilon]^3}
    F_\pm(p).
\end{align}
The resulting integers are the Chern numbers of the upper and lower projected eigenbundles on the enclosing surface. For an isolated two-band Weyl degeneracy, they satisfy $C_+=-C_-$.

Although the original FHS formulation was introduced for a discretized Brillouin zone~\cite{Fukui2005}, the same lattice-gauge construction applies to any discretized closed two-dimensional surface supporting a smooth nondegenerate eigenbundle. Here, the surface is a small cube in local wavevector space, rather than a Brillouin zone. We use Cartesian coordinates $(\delta k_x,\delta k_y,\delta k_z)$ because $k_\perp=\sqrt{k_x^2+k_y^2}$ is not a signed smooth coordinate at $k_\perp=0$. The plasma frequency is held fixed within each homogeneous calculation; changing $\omega_p$ corresponds to repeating the same local computation for a different homogeneous background.

In the numerical implementation, the computed charge is accepted only when the enclosed crossing is isolated, the projected eigenvalues remain nondegenerate on the entire cube boundary, and the integer is stable under modest changes in $\epsilon$ and the face resolution.

\bibliographystyle{apsrev4-2}
\bibliography{ref}

\end{document}